\documentclass[
 prl,
 groupedaddress,
 twocolumn,
 showpacs,
 preprintnumbers,
 amsmath,
 amssymb]
{revtex4}
\usepackage{amssymb}
\usepackage{graphicx}
\usepackage{dcolumn}
\usepackage{bm}
\usepackage{amsmath}

\begin{document}

\title{Detection of mechanical resonance of a single-electron transistor by direct current}

\author{Yu.~A.~Pashkin$^1$}
\email{pashkin@zp.jp.nec.com}
\altaffiliation[on leave from ]{P. N. Lebedev Physical Institute, Moscow 119991,
Russia}
\author{T.~F.~Li$^{1,2}$}
\author{J.~P.~Pekola$^{3}$}
\author{O.~Astafiev$^{1}$}
\author{D.~A.~Knyazev$^{4}$}
\author{F.~Hoehne$^{5}$}
\author{H.~Im$^{1,6}$}
\author{Y.~Nakamura$^{1}$}
\author{J.~S.~Tsai$^{1}$}
\affiliation{$^{1}$NEC Nano Electronics Research Laboratories and RIKEN Advanced Science Institute, Tsukuba,
Ibaraki 305-8501, Japan\\
$^{2}$Institute of Microelectronics, Tsinghua University, Beijing 100084, China\\
$^{3}$Low Temperature Laboratory, Aalto University School of Science and Technology, P.O. Box 13500, FI-00076
AALTO, Finland\\
$^{4}$P.~N.~Lebedev Physical Institute, Russian Academy of Sciences, Moscow 119991, Russia\\
$^{5}$Walter Schottky Institut, Technische Universit\"{a}t M\"{u}nchen, Am Coulombwall 3, 85748 Garching,
Germany\\
$^{6}$Department of Semiconductor Science, Dongguk University, Phil-Dong, Seoul 100-715, Korea}

\begin{abstract}

\noindent We have suspended an Al based single-electron transistor whose island can resonate freely between
the source and drain leads forming the clamps. In addition to the regular side gate, a bottom gate with a
larger capacitance to the SET island is placed underneath to increase the SET coupling to mechanical motion.
The device can be considered as a doubly clamped Al beam that can transduce mechanical vibrations into
variations of the SET current. Our simulations based on the orthodox model, with the SET parameters estimated
from the experiment, reproduce the observed transport characteristics in detail.
\end{abstract}

\maketitle

Nanomechanical resonators, usually doubly clamped beams or cantilevers, offer rich physics as well as a wide
range of applications~\cite{Ekinci,SchwabRoukes}. In order to do experiments on them, one needs a detector,
called a transducer, coupled to the resonator, which converts mechanical displacement into an electrical
signal. A number of techniques have been applied to measure mechanical motion at the micro and nanoscales.
For the past decade in the quest for higher sensitivity and speed, the dimensions of the resonators were
scaled down, pushing their resonance frequency to above 1 GHz~\cite{Huang}. At the same time the requirements
to the transducers become more stringent in terms of sensitivity to the mechanical displacement, thus
narrowing the choice of possible detectors. Among various detection techniques described
elsewhere~\cite{Blencowe}, a single-electron transistor (SET)~\cite{AverinLikh86,Fulton} proved to be an
efficient transducer due to its extremely high sensitivity and a capability of detecting the motion of a
mechanical resonator in the quantum limit. When capacitively coupled to the resonator and biased at a dc
voltage, the SET senses the resonator's mechanical motion due to the variations of the electrical charge
induced on the SET island. Using a radiofrequency (rf) circuitry and an SET as a mixer, displacement
sensitivity as good as $2\times10^{-15}$~m/Hz$^{1/2}$ was achieved for a GaAs mechanical
resonator~\cite{Knobel}. In the later experiment, an rf version of the SET was used to detect the thermal
motion of the SiN resonator demonstrating position resolution only a factor of 4.3 above the quantum
limit~\cite{Schwab}.
\newline
\indent In this letter, we describe an aluminum structure that combines both the doubly clamped beam and SET,
and therefore can be referred to as a two-in-one device. Our fabrication process allows easy integration of
metallic nanomechanical resonators into the electronic circuits such as SETs or SQUIDs. We show
experimentally that a conventional SET in the dc regime can detect flexural motion of its own island. We
observe the frequency response of the suspended SET driven by an rf voltage applied to the bottom gate. The
results are reproduced in the simulations based on the orthodox theory that takes into account the mechanical
degree of freedom of the transistor. A similar device based on a carbon nanotube is reported in
Refs.~\cite{Delft2009,Lassagne}.
\newline
\indent The device as well as wiring and voltage sources are shown schematically in Fig.1. The central part
of the SET is suspended above the substrate. It consists of an island connected to the source and drain
electrodes through Al/AlO$_{x}$/Al tunnel junctions. There is a vacuum gap between the island as well as
partly the source and drain electrodes and the bottom gate. Part of the island (shaded in Fig.1) clamped
between the source and drain electrodes can resonate. The suspension is made using the fabrication process
described in Ref.~\cite{LiAPL2008}.
\begin{figure}[bp]
\centering
\includegraphics[width=0.45\textwidth]{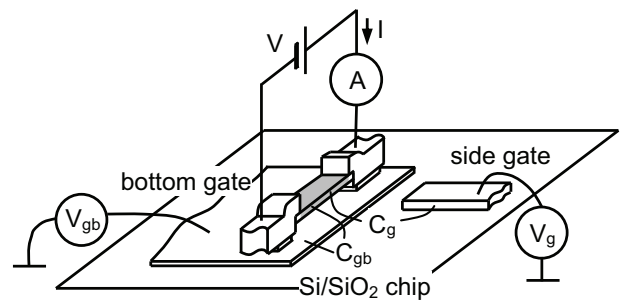} 
\caption{\protect\small {Schematics of the suspended SET and measurement circuit.}}
\end{figure}
However, here we introduced one important modification. In addition to
the regular side gate, an extra control electrode, called bottom gate, is placed underneath the island and
partly under the source and drain electrodes. Such a multi-layer two-gate configuration has certain
advantages over the standard one-layer layout. First, the coupling between the bottom gate and the island can
be made several times larger as compared to the one-side gate configuration implemented in a single layer.
This makes the SET more sensitive to mechanical motion. Second, the gap between the bottom gate and the
island depends on the thickness of the corresponding polymer used as a sacrificial layer and therefore can be
controlled accurately. Third, a high dc voltage $V_{dc}$ and slowly varying voltage $V_{g}$ can be applied to
different gates simplifying the measurement process. The voltage applied to the bottom gate has two
components: $V_{gb} = V_{dc} + V_{rf}$. The former is used to control coupling between the mechanical motion
and SET transport while the latter drives the beam. All the dc voltages are supplied to the sample by the
filtered dc wires. The rf signal is delivered through a coaxial line with a 20~dB attenuator at 4.2~K. The
measurements are done in a dilution refrigerator with a mixing chamber temperature of about 25~mK.

To model SET transport in the presence of mechanical oscillations, we perform simulations based on the
orthodox theory~\cite{AverinLikh} with the mechanical degree of freedom taken into account. We consider the
classical dynamics of the SET island and solve the equation of motion for the coordinate $x$, which is the
displacement of the beam center from the equilibrium position:

\begin{equation}
\ddot{x}+\frac{\omega_{0}}{Q}\dot{x}+\omega_{0}^{2}x=F/m,
\end{equation}

\noindent where $F$ is the driving force acting on the SET island, $m$ is the beam effective mass, $Q$ is the
quality factor and $\omega_0= 2\pi f_{0}$ is the angular resonance frequency. The force can be found as a
displacement derivative of the total energy stored in the SET island. Assuming $C_{gb} \ll C$, where $C$ is
the total capacitance of the SET island, we obtain $F \simeq \displaystyle\frac{\partial C_{gb}}{\partial
x}[\displaystyle\frac{1}{2}V_{gb}^{2}-\frac{e V_{gb}}{C}(n+\frac{C_{gb}V_{gb}}{e})]$, where $n$ is the
instant number of electrons on the SET island. We then further assume that the tunneling of electrons is much
faster than the mechanical oscillations, so that the tunneling rate for each event can be calculated for
constant $x$. The rates are calculated using the golden rule approach assuming both SET junctions are equal.

We first characterize the transistor by measuring the SET current $I$ as a function of the bias voltage $V$
and two gate voltages: side gate voltage $V_{g}$ and bottom gate voltage $V_{dc}$. From these measurements we
estimate the following parameters of the device: total tunnel resistance $R$ = 140~k$\Omega$, charging energy
$E_c = e^{2}/2C = 0.235$~meV ($E_{c}/k_{B} = 2.7$~K) corresponding to $C = 3.4\times10^{-16}$~F, and side
gate capacitance $C_{g} = 1.2\times10^{-18}$~F. By sweeping the voltage applied to the bottom gate, we obtain
its capacitance $C_{gb} = 5.4\times10^{-17}$~F. This value is in agreement with the naive estimation
$2.4\times10^{-17}$~F from the parallel-plate geometry, which underestimates the capacitance due to not
taking into account the fringing effects as well as additional bottom gate - island coupling through the
tunnel junctions. In this estimation, the island width $w = 92$~nm and length 1500~nm, and the vacuum
($\varepsilon = 1$) gap $d = 50$~nm between the island and bottom gate, measured in the scanning electron
microscope, were used. With the given dimensions of the  $38.6$~nm-thick island, and Al material parameters,
such as the mass density 2700~kg/m$^{3}$ and Young's modulus $70$~GPa, we estimate the unstressed beam
resonance frequency for the out-of-plane fundamental flexural mode to be about 90~MHz. However, as expected,
the resonance frequency measured in the experiment is higher due to the tensile stress produced by the
difference in thermal expansion coefficients of Al and Si \cite{LiAPL2008}.

In order to detect the beam resonance, we drive the beam with an external force by applying an rf voltage to
the bottom gate inducing out-of-plane oscillations. To increase the coupling of the SET to the mechanical
oscillations, we simultaneously apply to the same gate a high, up to $\pm 4$~V, dc voltage. The search for
the resonance is done in such a way that at constant rf amplitude and frequency, we sweep $V_g$ and measure
current $I$ through the SET. Then the frequency is increased by an increment of 3~kHz, which is smaller than
the expected resonance width $f_{0}/Q$. Even in the absence of mechanical resonance the modulation peak gets
slightly suppressed and broadened due to the fact that the working point shifts periodically when $V_{rf}$ is
applied (see Fig.~2(a)).
\begin{figure}[bp]
\centering
\includegraphics[width=0.45\textwidth]{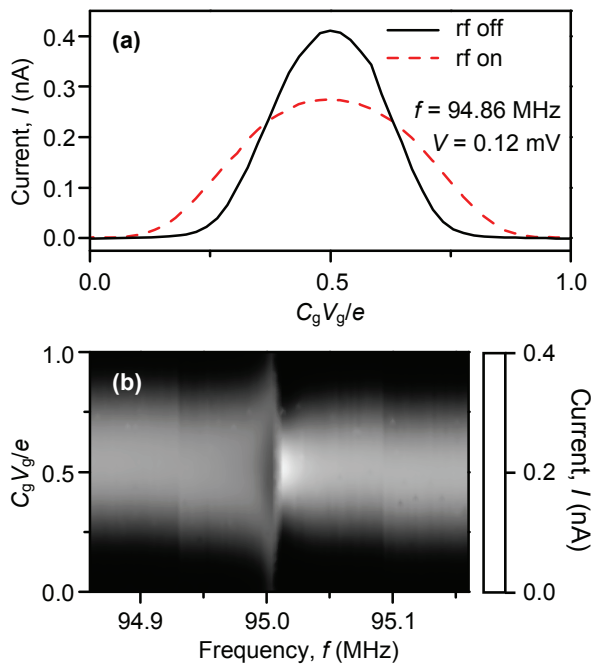} 
\caption{{\protect\small {(Color online) (a) $I$ vs $V_{g}$ curves for $V_{dc}=-2.5~V$ and the amplitude of
$V_{rf}$ equal to 0 (black curve) and 0.32~m$V$ (red curve) (b) Intensity plot showing an SET modulation peak
for the same values of $V_{dc}$ and the amplitude of $V_{rf}$ when the frequency of the applied drive is
varied around resonance.}}}
\end{figure}
\begin{figure}[tbp]
\centering
\includegraphics[width=0.4\textwidth]{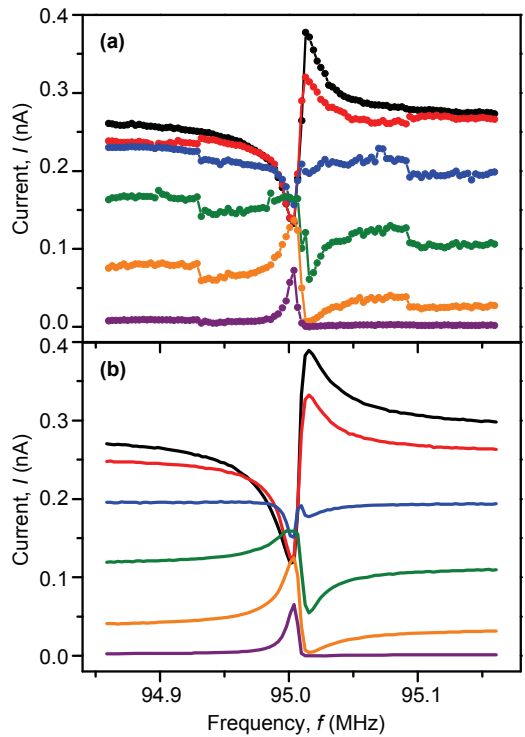}
\caption{{\protect\small {(Color online) Frequency dependence of the SET current at $V_{dc}=-2.5~V$,
amplitude 0.32~mV of $V_{rf}$ and at several values of $C_{g} V_{g}/e$. (a) Response curves measured at
$C_{g} V_{g}/e$ = 0.5, 0.4, 0.58, 0.77, 0.85 and 0.95 (from top to bottom); (b) Corresponding simulated
curves.}}}
\end{figure}
This effect, however, does not depend on frequency. Once the driving frequency
approaches the resonance frequency of the beam, there is an additional effect on the modulation peak, which
strongly depends on the frequency. The intensity plot revealing the expected resonance is presented in
Fig.~2(b). The resonance is seen as a characteristic feature at about 95~MHz being most pronounced at
$C_gV_g/e=0.5$: the current peak height decreases first and then suddenly increases when we go through
resonance from lower to higher frequency.

A collection of the SET response curves measured at various $V_g$ and the corresponding simulated curves are
shown in Fig.~3. In the simulations, we set $Q=10^4$ and $f_0=95.007$~MHz; other parameters were taken from
the experiment. The simulations capture all the essential features observed in the experiment. The current
steps at about 94.93~MHz and 95.09~MHz in Fig. 3(a) are due to the jumps of the background charge, which were
not accounted for in the simulations. Note that the observed response is expected in the fully linear regime;
no non-linearity of the resonator was included in the model. The observed dispersive-like resonance curve at
$C_gV_g/e=0.5$ instead of a Lorentzian expected for the driven harmonic oscillator in the linear regime can
be qualitatively understood in the following way. When we approach the resonance from the lower frequency
side, the SET modulation peak becomes more suppressed and smeared because both the driving force and the
displacement effect act in phase. When we pass the resonance, the displacement and the driving force become
shifted by 180 degrees with respect to each other. Therefore they compensate each other, and the SET current
rises to almost its original value when no rf voltage was applied. At a higher frequency, the amplitude of
the mechanical oscillations decreases and the modulation peak becomes almost equal to the one measured at a
lower frequency.

The SET sensitivity to the mechanical motion can be estimated by assuming that the change in the island
charge produced by the mechanical displacement should be equal to the SET charge equivalent noise. Though the
charge noise was not measured in the device presented, we conservatively set the noise level at low
frequencies, where the 1/$f$ noise dominates, to $q_{n} = 10^{-3}$ $e/\rm{Hz}^{1/2}$ at 1~Hz, which is
typical for metallic SETs \cite{PTB}. The charge variations due to the beam displacement are $q_m = V_{dc}
\Delta C_{gb} \cong V_{dc} C_{gb}x/d$ assuming $x\ll d$. Thus, the displacement sensitivity is equal to
$\delta x = q_{n}d/V_{dc}C_{gb}\approx2\times10^{-13}$ m/Hz$^{1/2}$ per volt of $V_{dc}$, which, in our
low-frequency measurement, corresponds to the $rms$ displacement amplitude $\sim10^{-12}$m. However, when the
SET in the dc regime measures high-frequency mechanical oscillations, the conversion of the beam displacement
into the charge variation of the SET island is a result of rectification of the rf signal on the SET current
nonlinearity, which gives an even lower displacement sensitivity. On the other hand, in the MHz frequency
range the SET charge noise is about two orders of magnitude lower~\cite{Wahlgren}. Therefore, the rf type SET
is able to resolve thermal motion of the nanomechanical resonator even at the mK temperature~\cite{Schwab}.
The $rms$ amplitude of the beam main flexural mode due to thermal fluctuations is estimated from the
equipartition theorem according to the formula $\langle{x_{T}^2}\rangle^{1/2} =
(k_{B}T/m\omega_{0}^{2})^{1/2}$, where $T$ is the temperature. This gives the value
$\langle{x_{T}^2}\rangle^{1/2} = 2.5\times10^{-13}$~m, which is not detectable in the present setup using dc
measurement.

The reduced displacement sensitivity of the device in the dc regime, as compared to the rf regime
($\sim10^{-15}$ m/Hz$^{1/2}$), by no means compromises performance of the SET as a detector but just a result
of simplification of the measurement circuit. Besides detecting its own mechanical resonance, the device
described can also be used for spectroscopy measurement of a suspended charge qubit. Another challenging
experiment is the observation of the lasing effect in the circuit containing an artificial atom (charge
qubit) coupled to a high-frequency mechanical resonator instead of a commonly used optical or microwave
resonator.
\newline
\indent We thank S. Asshab, N. Lambert and F. Nori for fruitful discussions. This work was supported by
CREST-JST, MEXT kakenhi ``Quantum Cybernetics" and the Academy of Finland.

\end{document}